\documentclass[11pt]{article}
\usepackage{amsmath, graphicx,setspace} \usepackage[margin=2.2cm]{geometry}
\title{Mapping Approach for Quantum-Classical Time Correlation Functions}
\author{Ali Nassimi$^a$ and Raymond Kapral$^b$\\ Chemical Physics Theory Group, Department of Chemistry, \\
University of Toronto, Toronto, ON, M5S 3H6, Canada\\ $^a$Email: anassimi@chem.utoronto.ca, Telephone: 416-946-7849  Fax: 416-978-5325\\$^b$Email: rkapral@chem.utoronto.ca, Telephone: 416-978-6106  Fax: 416-978-5325}
\date{}
\usepackage{wrapfig}
\begin{document}
\maketitle
\bibliographystyle{unsrt}
\begin{abstract}
The calculation of quantum canonical time correlation functions is
considered in this paper. Transport properties, such as diffusion
and reaction rate coefficients, can be determined from time
integrals of these correlation functions. Approximate,
quantum-classical expressions for correlation functions, which are
amenable to simulation, are derived. These expressions incorporate
the full quantum equilibrium structure of the system but
approximate the dynamics by quantum-classical evolution where a
quantum subsystem is coupled to a classical environment. The main
feature of the formulation is the use of a mapping basis
where the subsystem quantum states are represented by fictitious
harmonic oscillator states. This leads to a full phase space
representation of the dynamics that can be simulated without
appeal to surface-hopping methods. The results in this paper form
the basis for new simulation algorithms for the computation of
quantum transport properties of large many-body systems.
\subsection*{keywords} Mapping basis, Quantum-classical dynamics, Quantum correlation functions
\end{abstract}

\section{Introduction}
In the investigation of condensed phase systems one is usually
interested in the average value of some observable or in a time
correlation function from which a transport coefficient can be
computed. Since the basic description of matter is quantum
mechanical, we are interested in the average value of a quantum
mechanical operator, which is given by $\overline{{B}(t)} = {\rm
Tr} \hat{B}(t) \hat{\rho}(0)$, where, for a system with
Hamiltonian $\hat{H}$, the time dependent operator satisfies the
Heisenberg equation of motion,
\begin{equation} \label{quantum-Liouville}
\frac{d}{d t}\hat{B}(t)=\frac{i}{\hbar}[\hat{H},\hat{B}],
\end{equation}
and $\hat{\rho}(0)$ is the initial value of the density matrix.
Quantum time correlation functions of two operators, $\hat{A}$ and
$\hat{B}$ have the form\footnote{Other forms of quantum correlation functions
are useful in applications. These include symmetrized and Kubo transformed forms.
Since relations exist among these correlation functions~\cite{kubo}, we restrict
our considerations to this expression.}, $C_{AB}(t)=\rm{Tr}(\hat{\rho}_{eq}
\hat{A}\hat{B}(t))$, where $\hat{\rho}_{eq}$ is the quantum
canonical equilibrium density matrix. Either of these quantum
expressions requires a knowledge of the time evolution of a
quantum operator in a many-body system that is often very large.
Consequently, these general expressions are not computationally
tractable and appeal must be made to approximations if they are to
be evaluated for problems of physical interest.

The approximate dynamical description we consider in this paper is
quantum-classical Liouville dynamics~\cite{rayrev06}. In this formulation the
system is partitioned into two subsystems, which we call
quantum subsystem and bath or environment. The partition is
dictated by physical principles. For example, in electron or
proton transfer problems, the electron or proton may constitute
the quantum subsystem while the environment in which the transfer
takes place, a molecular group or biomolecule dissolved in a
solvent, forms the bath. In quantum-classical Liouville dynamics
the bath is treated classically while retaining the full quantum
character of the quantum subsystem. In this theory the analog of
the Heisenberg equation of motion for an operator is~\cite{rayrev06,ray99}
\begin{equation} \label{quantum-classical}
\frac{d}{d t}\hat{B}_W(X,t)=\frac{i}{\hbar}[\hat{H}_W,\hat{B}_W]-\frac{1}{2}
(\{\hat{H}_W,\hat{B}_W\}-\{\hat{B}_W,\hat{H}_W\}),
\end{equation}
where $X=(R,P)$ are the positions and momenta of the bath degrees
of freedom. The subscript W indicates that operators are expressed
in a partial Wigner representation defined below. In addition to
the usual quantum commutator, the equation of motion also involves
a Poissson bracket denoted by $\{\cdot,\cdot \}$. While this
equation is far more tractable than the full quantum equations of
motion, its numerical simulation presents challenges and a number
of different schemes have been constructed for this purpose~\cite{RAR,Sergi:MacKernan}. These
include Trotter-based methods that use an adiabatic basis~\cite{mck}, as well
as trajectory methods that use the diabatic~\cite{donmar} and force bases~\cite{ws00},
schemes based on the multiple threads algorithm~\cite{ws02} and a method that
utilizes the mapping basis~\cite{hyo:ray:ali}. The mapping method has
proved to be especially promising for the evaluation of
expectation values of operators, as shown by calculations of
population relaxation in the spin-boson model, one of the standard
test cases for quantum dynamics~\cite{hyo:ray:ali}. In this paper, we show how the
mapping method can be extended to the computation of
time correlation functions within the quantum-classical Liouville
framework.

In the next section, we present the explicit expression for the
quantum correlation function that forms the basis of our
calculations. Section~\ref{sec:mapping} outlines the mapping
formulation where subsystem quantum states are
replaced by fictitious harmonic oscillator states.
Section~\ref{sec:map-corr-function} reformulates the quantum
correlation function in the mapping basis. This fully
quantum description is exact and the dynamics is embodied in the
spectral density function. A quantum-classical approximation for
the spectral density dynamics is derived in
Sec.~\ref{sec:W-dynamics}. The details of this derivation are
presented in the Appendix. This result is used in
Sec.~\ref{sec:QC-corr-function} to obtain the final result for the
quantum-classical correlation function in the mapping basis. The conclusions of the study are presented in
Sec.~\ref{sec:conc}.

\section{Correlation Function}\label{sec:corr-function}

Before considering quantum-classical approximations to the
dynamics, we first rewrite the exact quantum correlation function
in a form that is suitable for the introduction of the mapping
basis and passage to the quantum-classical limit. Suppose the
quantum subsystem and bath have $N_s$ and $N_b$ degrees of
freedom with characteristic masses $m$ and $M$, respectively. In
order to introduce a phase space description of the bath, we first
introduce a coordinate representation of the bath so that
the correlation function takes the form,
\begin{eqnarray}\label{correlation}
C_{AB}(t)&=&\rm{Tr}(\hat{\rho}_{eq}
\hat{A}\hat{B}(t))=\rm{Tr}(\hat{\rho}_{eq} \hat{A}e^{i\hat{H}t/\hbar}\hat{B}e^{-i\hat{H}t/\hbar}) \nonumber\\
&=& \rm{Tr}' \int dQ_1dQ_2dQ_3dQ_4 \langle Q_1|\hat{\rho}_{eq}\hat{A}|Q_2
 \rangle  \langle Q_2|e^{i\hat{H}t/\hbar}|Q_3 \rangle \langle Q_3|\hat{B}|Q_4
 \rangle  \langle Q_4|e^{-i\hat{H}t/\hbar}|Q_1 \rangle  \nonumber \\
&=&  \rm{Tr}' \int dR_1dR_2dZ_1dZ_2 \langle R_1-\frac{Z_1}{2}|\hat{\rho}_{eq}\hat{A}|R_1+\frac{Z_1}{2}
 \rangle  \langle R_1+\frac{Z_1}{2}|e^{i\hat{H}t/\hbar}|R_2-\frac{Z_2}{2} \rangle  \nonumber\\
&&\quad \times \langle R_2-\frac{Z_2}{2}|\hat{B}|R_2+\frac{Z_2}{2}
 \rangle  \langle R_2+\frac{Z_2}{2}|e^{-i\hat{H}t/\hbar}|R_1-\frac{Z_1}{2} \rangle.
\end{eqnarray}
In these equations, $\rm{Tr}'$ stands for a trace over the quantum
subsystem degrees of freedom and, in the forth equality above, we have
made a change of variables $Q_1=R_1-Z_1/2$, $Q_2=R_1+Z_1/2$, etc.

Two further manipulations are required to cast the correlation
function into a form that is suitable for the calculations using
the mapping basis described below. The coordinate
space matrix elements may be replaced with bath phase space
functions by introducing the partial~\cite{ray99} Wigner transforms~\cite{wigner,hillery,imre} of an
operator and density matrix,
\begin{eqnarray}\label{Weil}
\langle R-\frac{Z}{2}|\hat{B}|R+\frac{Z}{2}\rangle &=&
\frac{1}{(2\pi\hbar)^{N_b}} \int dP \hat{B}_W(R,P) e^{-iP \cdot
Z/\hbar}, \nonumber \\
\langle R-\frac{Z}{2}|\hat{\rho}_{eq}\hat A|R+\frac{Z}{2}\rangle &=&
\int dP (\hat{\rho}_{eq}\hat A)_W(R,P) e^{-iP \cdot Z/\hbar}.
\end{eqnarray}
Note that the partially Wigner transformed quantities are still operators in
the quantum subsystem Hilbert space. In addition, ${\rm Tr}'$ may be written
explicitly using the eigenfunctions of the Hamiltonian of the
quantum subsystem. The Hamiltonian for the entire system is given
by
\begin{equation}
\label{hamiltonian} \hat H = \frac{\hat P^2}{2M} + \hat V_b(\hat
R) + \frac{\hat p^2}{2m} + \hat V_s(\hat q) + \hat V_c(\hat R,\hat
q),
\end{equation}
where $\hat P$ and $\hat p$ are momentum operators of the bath and
subsystem and $\hat V_b$, $\hat V_s$ and $\hat V_c$ are, respectively, the bath,
subsystem and coupling potentials. The coordinate operators for
the subsystem and bath are $\hat{q}$ and $\hat{R}$, respectively.
Equation~(\ref{hamiltonian}) can be written as $\hat H =
\frac{\hat P^2}{2M}+\hat V_b+\hat V_c+\hat{h}_s$, where
$\hat{h}_s=\frac{\hat p^2}{2m}+\hat V_s$ is the subsystem
Hamiltonian. The eigenstates of $\hat{h}_s$ are defined by the
eigenvalue problem $\hat{h}_s|\lambda\rangle =
\epsilon_{\lambda}|\lambda\rangle$. We suppose that there are $N$ quantum
subsystem states. Making use of subsystem energy eigenstates as a basis and
introducing the Wigner transformed forms of $\hat \rho_{eq} \hat A$ and $\hat B$
given in Eq.~(\ref{Weil}) we have,
\begin{eqnarray}
\label{correlation2} C_{AB}(t)&=& \frac{1}{(2\pi\hbar)^{N_b}} \sum_{\lambda, \lambda', \mu, \mu'=1}^{N}
\int dX_1dX_2dZ_1dZ_2\; \langle \lambda|(\hat{\rho}_{eq}\hat{A})_W(X_1)|{\lambda'} \rangle \nonumber \\
&&\times \langle {\lambda'}|\langle
R_1+\frac{Z_1}{2}|e^{i\hat{H}t/\hbar}|R_2-\frac{Z_2}{2}
\rangle |\mu \rangle  \langle \mu|\hat{B}_W(X_2)|{\mu'} \rangle \nonumber \\
&&\times \langle {\mu'}|\langle
R_2+\frac{Z_2}{2}|e^{-i\hat{H}t/\hbar}|R_1-\frac{Z_1}{2} \rangle
|\lambda \rangle e^{-\frac{i}{\hbar}(P_1 \cdot Z_1 + P_2 \cdot Z_2)}.
\end{eqnarray}
The spectral density function in the subsystem basis may be defined as
\begin{eqnarray}
W^{\lambda' \lambda \mu' \mu}(X_1,X_2,t)&=&  \int dZ_1dZ_2 \;
\langle {\lambda'}|\langle
R_1+\frac{Z_1}{2}|e^{i\hat{H}t/\hbar}|R_2-\frac{Z_2}{2} \rangle
|\mu \rangle \nonumber \\
&& \times \langle {\mu'}|\langle
R_2+\frac{Z_2}{2}|e^{-i\hat{H}t/\hbar}|R_1-\frac{Z_1}{2} \rangle
|\lambda \rangle e^{-\frac{i}{\hbar}(P_1 \cdot Z_1 + P_2 \cdot Z_2)},
\end{eqnarray}
and contains all information needed to compute the quantum time
evolution contribution to the correlation function. The spectral density defined via full Wigner transform was used by Filinov \emph{et al.} \cite{Fil95,Fil96,Fil96b}
in a reformulation of the quantum correlation function. In terms of
the spectral density function in the subsystem basis the correlation function
takes the form,
\begin{equation}
\label{correlation2a} C_{AB}(t)=
\frac{1}{(2\pi\hbar)^{N_b}}\sum_{\lambda, \lambda',
\mu,\mu'}^{N} \int dX_1dX_2 \; \langle
\lambda|(\hat{\rho}_{eq}\hat{A})_W(X_1)|{\lambda'} \rangle
\langle \mu|\hat{B}_W(X_2)|{\mu'} \rangle W^{\lambda'\lambda \mu'
\mu}(X_1,X_2,t).
\end{equation}

This expression for the quantum correlation function is exact. In
order to compute it, the matrix elements of the forward and
backward propagators must be evaluated to solve for the time
dependence of the spectral density function. This is the most
difficult part of the problem. A similar expression utilizing the
adiabatic basis in place of the subsystem basis was derived
earlier~\cite{Rayser04}. In addition to the time dependent spectral density
function, the time independent matrix elements of the quantum
operators and quantum equilibrium density matrix must also be computed
to evaluate the ensemble average appearing in the definition of
the correlation function. While these equilibrium quantities may
be difficult to evaluate for complex systems, they are far easier
to compute than the quantum time dependence, and algorithms have
been developed for this purpose~\cite{rossky,geva}.

\section{Mapping Basis} \label{sec:mapping}

In order to construct a useful simulation algorithm for the quantum
correlation function, it is convenient to use an alternative, but equivalent,
mapping form for the quantum subsystem matrix elements of the
operators which enter in its definition. A well-known mapping
approach was introduced by Schwinger~\cite{sch}. In his scheme, the eigenstates of
the angular momentum operator are mapped onto eigenfunctions of two bosonic oscillators,
and the angular momentum operators are mapped onto combinations of
creation, $\hat{a}_{\lambda}^\dag$, and annihilation,
$\hat{a}_{\lambda'}$, operators ($\lambda, \lambda' =1$ or 2). Such a mapping
yields a simple treatment of the angular momentum problem in quantum
mechanics. In this formalism the resolution of identity
is mapped to
$\hat{a}_{1}^\dag \hat{a}_{1}+\hat{a}_{2}^\dag \hat{a}_{2}=1$. This
equality is used in the Holstein-Primakoff mapping scheme to
eliminate one bosonic oscillator and represent angular momentum
states by a single oscillator~\cite{holpri40}. An extension of
these mapping schemes was used to map discreet states of a quantum system
onto fictitious harmonic oscillators, so that all degrees of
freedom in the system can be treated with semiclassical
approximations \cite{StoTho97, MulSto98, ThoSto99, stotho, Mill:Mcur, sun98, Miller01}. The mapping
approach has also been used in other closely related contexts
\cite{bonella03, bonella05, Dun:Bon:Cok}.

In the mapping scheme used in this work, an $N$ level quantum
system is mapped onto $N$ harmonic oscillators. The wavefunction
of the system in a given quantum state is mapped onto a product of
harmonic oscillator wavefunctions where all oscillators are in their
ground state, except for one oscillator corresponding to the given
quantum state, which is in its first excited state. This mapping is schematically represented in Fig~\ref{fig1}.  Therefore, we
have a physical space with a cardinality $N$, which is much lower
than the cardinality of the Hilbert space of the $N$ harmonic
oscillators, which is infinite.
More specifically, the subsystem quantum states are mapped through the relations
\begin{equation}\label{mapping1}
|\lambda \rangle \rightarrow |m_\lambda \rangle
=|0_1,\cdots,1_\lambda,\cdots,0_N \rangle ,
\end{equation}
where
\begin{equation}
\langle q|m_{\lambda}\rangle = \langle
q_{1},q_{2},\cdots,q_{N}|0_{1},\cdots,1_{\lambda},\cdots,0_{N}\rangle \nonumber =
\phi_{0}(q_{1})\cdots\phi_{0}(q_{\lambda-1})\phi_{1}(q_{\lambda})\cdots\phi_{0}(q_{N})\label{coordinatereporesentationofmapping},
\end{equation}
with $\phi_{0}$ and $\phi_{1}$, respectively, being the ground and
the first excited state wavefunctions of an harmonic oscillator.
The creation and annihilation operators on the mapping states act
in the following ways\footnote{Since the harmonic oscillators are
fictitious we may set $m\omega$ to 1 to simplify these
expressions; however,  we retain these forms both to make
dimensional consistency manifest and to simplify the dynamical
relations (see Sec.~\ref{sec:W-dynamics}) by equating m to the
quantum subsystem characteristic mass and $\omega$ to the inverse
of the scaling time.}:
\begin{figure}[tb]
\setlength{\unitlength}{1cm}
\begin{center}
\begin{picture}(10,5.5)
\multiput(.5,1)(0,.8){5}{\line(1,0){1}}
\put(1,3.4){\circle*{.15}} \put(1.9,2.6){\vector(1,0){1.4}}
\multiput(3.8,1)(0,.6){7}{\line(1,0){.7}}
\multiput(4.8,1)(0,.6){7}{\line(1,0){.7}}
\multiput(5.8,1)(0,.6){7}{\line(1,0){.7}}
\multiput(6.8,1)(0,.6){7}{\line(1,0){.7}}
\multiput(7.8,1)(0,.6){7}{\line(1,0){.7}}
\put(4.15,1){\circle*{.15}} \put(5.15,1){\circle*{.15}}
\put(6.15,1){\circle*{.15}} \put(7.15,1.6){\circle*{.15}}
\put(8.15,1){\circle*{.15}}
\put(3.4,0){\makebox(0,0){$|4\rangle \qquad \longrightarrow \qquad |0,0,0,1,0 \rangle $}}
\end{picture}
\end{center}

\caption{Schematic representation of the mapping for a 5 level
system. Subsystem states are on the left and the mapping states
are on the right.} \label{fig1}
\end{figure}
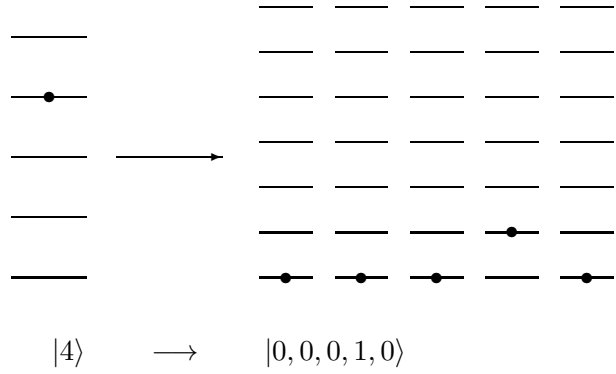
\begin{equation} \label{mapping}
\hat a_{\lambda}^\dag |0 \rangle =|0_1,\cdots,1_\lambda,\cdots,0_N \rangle
=|m_{\lambda} \rangle ,\qquad \text{and} \qquad \hat a_{\lambda}|m_{\lambda}
 \rangle =|0_1,\cdots,0_N \rangle  = |0 \rangle ,
\end{equation}
where
\begin{equation}
\label{creation} \hat{a}_{\lambda} = \sqrt{\frac{m \omega}{2
\hbar}}(\hat{q}_{\lambda}+\frac{i}{m \omega}\hat{p}_{\lambda}),
\hspace{0.6cm} \hat{a}_{\lambda}^{\dag}=\sqrt{\frac{m \omega}{2
\hbar}}(\hat{q}_{\lambda}-\frac{i}{m \omega}\hat{p}_{\lambda}),
\hspace{0.6cm} \text{and} \hspace{0.6cm}
[\hat{q}_{\lambda},\hat{p}_{\lambda}]=i \hbar.
\end{equation}
We may then introduce the mapping representation of an operator $\hat{A}$ as
\begin{equation}\label{eq:map-operator}
\hat A_m = \sum_{\lambda \lambda'} A_{\lambda \lambda'}
\hat{a}_{\lambda}^\dag \hat{a}_{\lambda'}.
\end{equation}
For example, the mapping form of the Hamiltonian in Eq.~(\ref{hamiltonian}) is
\begin{eqnarray}
\label{hamiltonian2} \hat{H}_m &=& \sum_{\lambda \lambda'} \big[\frac{\hat P^2}{2M}\delta_{\lambda\lambda'} + \hat
V_B(\hat R)\delta_{\lambda\lambda'} + h_{\lambda \lambda'}(\hat R) \big]
\hat a_{\lambda}^\dag \hat a_{\lambda'} \nonumber \\
&=& \frac{\hat P^2}{2M} + \hat V_B(\hat R) +
\frac{m\omega}{2\hbar} \sum_{\lambda \lambda'} h_{\lambda
\lambda'}(\hat R)\Big(\hat r_{\lambda} \hat r_{\lambda'} +
\frac{\hat p_{\lambda} \hat p_{\lambda'}}{m^2\omega^2} -
\frac{\hbar}{m\omega} \delta_{\lambda\lambda'}\Big),
\end{eqnarray}
where $\hat{h}=\frac{\hat p^2}{2m} + \hat V_s(\hat q) + \hat V_c(R,\hat q)$ and $h_{\lambda \lambda'}$ is its matrix element. We
used the fact that $h_{\lambda \lambda'} = h_{\lambda' \lambda}$
in writing this expression. Note that unlike
Eq.~(\ref{hamiltonian}) where $\hat q$ and $\hat p$ are the
subsystem coordinates and momenta, in Eq.~(\ref{hamiltonian2})
$\hat r$ and $\hat p$ are the mapping space coordinates and
momenta. These quantities have dimensions equal to the number
of subsystem quantum states.

From the definition in Eq.~(\ref{eq:map-operator}) it is clear
that the matrix elements of $\hat{A}$ in the subsystem basis are
identical to those of $\hat A_m$ in the mapping basis: $\langle
\lambda | \hat{A} |\lambda' \rangle =A_{\lambda \lambda'} =
\langle m_\lambda | \hat{A}_m |m_{\lambda'} \rangle$.
Consequently, any matrix element in the quantum subsystem basis
can be substituted by its equivalent form in the mapping basis. We
show that this substitution leads to computational advantages when
simulating quantum correlation functions.

\section{Correlation Function in Mapping Basis} \label{sec:map-corr-function}
The matrix elements in the correlation function
expression~(\ref{correlation2}) can be replaced by their mapping
equivalent forms to yield
\begin{eqnarray}
\label{correlation4}
C_{AB}(t)&=& \frac{1}{(2\pi\hbar)^{N_b}}
\sum_{\lambda\lambda' \mu \mu'=1}^{N} \int dX_1dX_2dZ_1dZ_2\;
\langle m_\lambda|(\hat{\rho}_{eq}\hat{A})_{Wm}(X_1)|m_{\lambda'} \rangle \nonumber \\
&&\times \langle m_{\lambda'}|\langle R_1+\frac{Z_1}{2}|
e^{i\hat{H}_mt/\hbar}|R_2-\frac{Z_2}{2} \rangle |m_\mu
\rangle  \langle m_\mu|\hat{B}_{Wm}(X_2)|m_{\mu'} \rangle \nonumber \\
&&\times \langle m_{\mu'}|\langle R_2+\frac{Z_2}{2}|
e^{-i\hat{H}_mt/\hbar}|R_1-\frac{Z_1}{2} \rangle |m_\lambda
\rangle e^{-\frac{i}{\hbar}(P_1 \cdot Z_1 + P_2 \cdot Z_2)}.
\end{eqnarray}
The next step in the analysis of this correlation function is to introduce
a coordinate space representation of the abstract mapping eigenfuctions.
Inserting resolutions of the identity, $\int dq \; |q\rangle \langle q|=1$ and
making use of the closure relation for mapping states, $\sum_\lambda \langle q'|m_\lambda \rangle
\langle m_\lambda |q \rangle=\delta(q-q')$, the correlation function takes the form
\begin{eqnarray} \label{correlation4a}
C_{AB}(t)&=& \frac{1}{(2\pi\hbar)^{N
_b}} \int dX_1dX_2dZ_1dZ_2 dq_1 dq_2 dq_4 dq_6\;
 \langle q_1|(\hat{\rho}_{eq}\hat{A})_{Wm}(X_1)|q_2 \rangle  \nonumber \\
&&\times \langle q_2|\langle R_1+\frac{Z_1}{2}|
e^{i\hat{H}_mt/\hbar}|R_2-\frac{Z_2}{2} \rangle |q_4 \rangle
\langle q_4|\hat{B}_{Wm}(X_2)|q_6 \rangle  \nonumber \\
&&\times \langle q_6|\langle
R_2+\frac{Z_2}{2}|e^{-i\hat{H}_mt/\hbar}|R_1-\frac{Z_1}{2} \rangle
|q_1 \rangle  e^{-\frac{i}{\hbar}(P_1 \cdot Z_1 + P_2 \cdot Z_2)}.
\end{eqnarray}
It is important to note that the dimensionality of the mapping coordinate
space representation is fixed by the number of quantum states.

A phase space description of this coordinate representation can be obtained by
introducing Wigner transforms of the matrix elements. Making the change of
variables, $q_1=r_1-z_1/2$, $q_2=r_1+z_1/2$, etc., and, using the analog of
Eq.~(\ref{Weil}) for the mapping coordinates, we find

\begin{eqnarray} \label{correlation5a}
C_{AB}(t) &=& \frac{1}{(2\pi\hbar)^{(N_b+N)}}
\int dX_1dX_2 dx_1 dx_2\; ({\hat \rho}_{eq}{\hat A})_{Wm}(X_1,x_1) \nonumber \\
&&\times {B}_{Wm}(X_2,x_2) W(X_1,X_2,x_1,x_2,t),
\end{eqnarray}
where, in analogy with the bath phase space terminology, $x=(r,p)$. The
full Wigner representation of the mapping spectral density, $W$, is defined as
\begin{multline}
\label{spectral density} W(X_1,X_2,x_1,x_2,t) = \int dZ_1dZ_2
dz_1 dz_2 \; \langle r_1+\frac{z_1}{2}|\langle R_1+\frac{Z_1}{2}|
e^{i\hat{H}_mt/\hbar}|R_2-\frac{Z_2}{2} \rangle |r_2-\frac{z_2}{2} \rangle \\ \times \langle r_2+\frac{z_2}{2}|\langle
R_2+\frac{Z_2}{2}|e^{-i\hat{H}_mt/\hbar}|R_1-\frac{Z_1}{2} \rangle
|r_1-\frac{z_1}{2} \rangle
e^{-\frac{i}{\hbar}(P_1 \cdot Z_1 + P_2 \cdot Z_2+p_1 \cdot z_1 + p_2 \cdot z_2)}.
\end{multline}
All of these manipulations have served simply to cast the exact expression for the
quantum correlation function into an equivalent exact form involving a phase-space-like
representation for both the bath and the mapping version of the quantum subsystem
degrees of freedom.  There are advantages to defining W and deriving its dynamics
rather than directly considering the dynamics of the correlation function. Not
only are the algebraic manipulations simplified, but also all
correlation functions for a specific system share the same spectral density.
This fully quantum problem is still intractable for complex
many-body systems. We now turn to the evaluation of the spectral density
function in the quantum-classical limit that will provide the basis for
simulation algorithms of the dynamics.

\section{Quantum-Classical Dynamics for $W$} \label{sec:W-dynamics}

Given the development presented above, the dynamical problem
consists in finding a tractable equation of motion for the
spectral density function $W(X_1,X_2,x_1,x_2,t)$ and constructing
an algorithm for its solution. The equation of motion can be
derived by differentiating $W$ with respect to time. However, to
derive the quantum-classical equation for $W$ we must introduce
approximations. For this purpose we make use of an expansion in
the small mass ratio $\mu =(m/M)^{1/2}$, where it is assumed that
the characteristic mass $M$ of the bath particles is much larger
than $m$ for the quantum subsystem particles. The manner in which
this expansion is carried out is analogous to that discussed
earlier for quantum Liouville equation~\cite{ray99}. More
specifically, given the energy $\epsilon_0$, time $t_0 =
\hbar/\epsilon_0$, and length $\lambda_m =
(\hbar^2/m\epsilon_0)^{1/2}$ units, we scale the light and heavy
particle momenta, respectively, with $p_m = (m \lambda_m/t_0) =
(m\epsilon_0)^{1/2}$ and $P_M = (M\epsilon_0)^{1/2}$. Note that
the only difference between subsystem and bath particles is
scaling their momenta with different factors. As the subsystem and
bath are in thermal equilibrium their average kinetic energies are
equal and therefore, on average, $p/P=\mu$, so that after scaling
both subsystem and bath momenta have the same order of magnitude.
After scaling variables, we obtain
\begin{multline}
\label{scaled spectral density}
W'(R'_1,P'_1,R'_2,P'_2,r'_1,p'_1,r'_2,p'_2;t') = \int dZ'_1dZ'_2 dz'_1 dz'_2 \langle r'_1+\frac{z'_1}{2}|\langle R'_1+\frac{Z'_1}{2}|e^{i\hat{H'}_mt'}|R'_2-\frac{Z'_2}{2} \rangle |r'_2-\frac{z'_2}{2} \rangle  \\
\times \langle r'_2+\frac{z'_2}{2}|\langle
R'_2+\frac{Z'_2}{2}|e^{-i\hat{H'}_mt'}|R'_1-\frac{Z'_1}{2} \rangle
|r'_1-\frac{z'_1}{2} \rangle
e^{-i\mu^{-1}(P'_1\cdot Z'_1 + P'_2\cdot Z'_2)}e^{-i(p'_1\cdot z'_1+p'_2\cdot z'_2)},
\end{multline}
where a prime means that the variable is divided by the
corresponding scaling factor. To avoid cumbersome notation, we
drop the primes in the following relations. Differentiation of the
scaled form of $W$ yields the equation of motion,
\begin{eqnarray}
\label{spectral density time derivative}
\frac{\partial W(t)}{\partial t} &=& i\int dZ_1 dZ_2 dz_1 dz_2
dQ dq \; \Big[\langle r_1+\frac{z_1}{2}|\langle R_1+\frac{Z_1}{2}|\hat{H}_m|Q \rangle |q \rangle  \nonumber \\
&&\times \langle q|\langle Q| e^{i\hat{H}_mt}|R_2-\frac{Z_2}{2} \rangle |r_2-\frac{z_2}{2}\rangle
\langle r_2+\frac{z_2}{2}|\langle R_2+\frac{Z_2}{2}|e^{-i\hat{H}_mt}|R_1-\frac{Z_1}{2}
\rangle |r_1-\frac{z_1}{2} \rangle \nonumber \\
&&-\langle r_1+\frac{z_1}{2}|\langle R_1+\frac{Z_1}{2}|e^{i\hat{H}_mt}|R_2-\frac{Z_2}{2}
\rangle |r_2-\frac{z_2}{2} \rangle  \langle r_2+\frac{z_2}{2}|\langle R_2+\frac{Z_2}{2}|
e^{-i\hat{H}_mt}|Q \rangle |q \rangle \nonumber \\
&&\times \langle Q|\langle q|\hat{H}_m|R_1-\frac{Z_1}{2} \rangle
|r_1-\frac{z_1}{2} \rangle \Big]
e^{-i\mu^{-1}(P_1 \cdot Z_1 + P_2 \cdot Z_2)}e^{-i(p_1 \cdot z_1 + p_2 \cdot z_2)}.
\end{eqnarray}
The mapping Hamiltonian (Eq.~(\ref{hamiltonian2})) in scaled
variables is $\hat{H}_m = \frac{\hat P^2}{2} + \hat V_B(\hat R) +
\sum_{\lambda \lambda'} \frac{h_{\lambda \lambda'}(\hat
R)}{2}(\hat r_{\lambda} \hat r_{\lambda'} + \hat p_{\lambda} \hat
p_{\lambda'} - \delta_{\lambda\lambda'})$, while the scaled
quantum mechanical momentum operators are
$\hat{P}=\frac{\mu}{i}\frac{\partial}{\partial Q}$ and $\hat{p} =
-i\frac{\partial}{\partial q}$. Substituting the Hamiltonian and
momentum operators into Eq.~(\ref{spectral density time
derivative}) yields,
\begin{eqnarray}
\label{spectral density time derivative2}
\frac{\partial W(t)}{\partial t} &=& i\mu^{2N_b}\int d\tilde{Z}_1 d\tilde{Z}_2
dQ dz_1 dz_2dq e^{-i(P_1\tilde{Z}_1+P_2\tilde{Z}_2)}e^{-i(p_1 \cdot z_1 + p_2 \cdot z_2)} \nonumber \\
&&\times \Bigg \{ \bigg\{ \Big[\frac{-\mu^2}{2}\frac{\partial^2}{\partial Q^2} + V_B(Q) +
\frac{1}{2}\sum_{\lambda \lambda'} h_{\lambda
\lambda'}(Q)(q_{\lambda}q_{\lambda'} - \frac{\partial}{\partial
q_{\lambda}} \frac{\partial}{\partial q_{\lambda'}}-\delta_{\lambda
\lambda'})\Big]\nonumber \\
&&\times \delta(Q-R_1-\frac{\mu \tilde{Z}_1}{2}) \delta(q-r_1-\frac{z_1}{2}) \bigg\} \langle q|\langle Q| e^{i\hat{H}_mt}|R_2-\frac{\mu \tilde{Z}_2}{2} \rangle |r_2-\frac{z_2}{2} \rangle \nonumber \\
&&\times \langle r_2+\frac{z_2}{2}|\langle R_2+\frac{\mu
\tilde{Z}_2}{2}|e^{-i\hat{H}_mt}|R_1-\frac{\mu \tilde{Z}_1}{2}
\rangle |r_1-\frac{z_1}{2} \rangle  \nonumber \\
&&- \bigg\{ \Big[\frac{-\mu^2}{2}\frac{\partial^2}{\partial Q^2} +
V_B(Q) + \frac{1}{2} \sum_{\lambda \lambda'} h_{\lambda
\lambda'}(Q)(q_{\lambda}q_{\lambda'} - \frac{\partial}{\partial
q_{\lambda}} \frac{\partial}{\partial
q_{\lambda'}}-\delta_{\lambda \lambda'})\Big]\nonumber \\
&& \times \delta(Q-R_1+\frac{\mu\tilde{Z}_1}{2}) \delta(q-r_1+\frac{z_1}{2}) \bigg\} \langle r_1+\frac{z_1}{2}|\langle R_1+\frac{\mu \tilde{Z}_1}{2}|e^{i\hat{H}_mt}|R_2-\frac{\mu \tilde{Z}_2}{2}
\rangle |r_2-\frac{z_2}{2} \rangle \nonumber  \\
&& \times \langle r_2+\frac{z_2}{2}|\langle R_2+\frac{\mu
\tilde{Z}_2}{2}|e^{-i\hat{H}_mt}|Q \rangle |q \rangle \Bigg \},
\end{eqnarray}
where the change of variable $\tilde{Z}_1 = Z_1/\mu$ was performed
in order to transfer the $\mu$ dependence from the exponential to
the argument of the potential energy terms, after integrations over $Q$ and
$q$ have been performed. It is more convenient to expand a potential energy
term around a small argument rather than to deal with an oscillatory exponential.

We begin the calculation by performing the integrals over $q$ and $Q$, followed by
Taylor expansion of the potential energy terms around $\mu=0$,
keeping terms up to the first order in $\mu$. Finally, we
introduce the definition of W into the resulting expression. The
algebra is lengthy so it is presented in the Appendix. While the algebra leading to
the result is lengthy, the final equation of motion is relatively simple. In
unscaled coordinates it takes the form,
\begin{eqnarray}
\label{spectral density time derivative5} \frac{\partial
W(t)}{\partial t} &=& - \frac{1}{\hbar} \sum_{\lambda \lambda'}
h_{\lambda \lambda'}(R_1)
\Big[r_{1\lambda}\frac{\partial}{\partial p_{1\lambda'}} -
p_{1\lambda} \frac{\partial}{\partial r_{1\lambda'}} \Big]W(t) +
(\frac{P_1}{M} \cdot \frac{\partial }{\partial R_1}
-\frac{\partial H_m}{\partial R_1} \cdot \frac{\partial }{\partial P_1})W(t)  \nonumber \\
&&+ \frac{\hbar}{8} \sum_{\lambda \lambda'} \frac{\partial
h_{\lambda \lambda'}(R_1)}{\partial R_1} \cdot
\Big(\frac{\partial}{\partial r_{1\lambda}}\frac{\partial}{\partial r_{1\lambda'}} +
\frac{\partial}{\partial p_{1\lambda'}}\frac{\partial}{\partial
p_{1\lambda}}\Big) \frac{\partial}{\partial P_1}W(t) \nonumber \\
&\equiv& i{\mathcal L}_m(x_1,X_1) W(t).
\end{eqnarray}
Here the Wigner transform of the Hamiltonian~(\ref{hamiltonian2})
is given by
\begin{equation}
\label{hamiltonian3} H_{m} = \frac{P^2}{2M} + V_B(R) +
\sum_{\lambda \lambda'} \frac{h_{\lambda \lambda'}(R)}{2\hbar}(
r_{\lambda}r_{\lambda'} + p_{\lambda}p_{\lambda'} - \hbar
\delta_{\lambda\lambda'}).
\end{equation}

This equation of motion is one of the principal results of this
paper. It results from an expansion of the evolution operator for
$W$ to order $\mu$ and is equivalent to quantum-classical
Liouville dynamics for the spectral density
function~\cite{Rayser04}. The first term in Eq.~(\ref{spectral
density time derivative5}) represents the subsystem dynamics of
the spectral density in the mapping phase space. The second term
is the dynamics of the spectral density due to classical evolution
of the bath degrees of freedom and the third term is a higher
order correlation between the quantum mapping and classical
degrees of freedom. The last equality defines the
quantum-classical Liouville operator in the mapping basis. This
equation must be solved subject to the initial condition,
\begin{eqnarray}
W(0)&=& (2 \pi \hbar)^{(N_b +N)} \delta(r_1-r_2) \delta(R_1-R_2)
\delta(p_1-p_2) \delta(P_1-P_2).
\end{eqnarray}

In Eq.~(\ref{spectral density time derivative}), since $\hat{H}_m$
commutes with propagator $e^{i\hat{H}_mt}$, when differentiating
with respect to time, $\hat{H}_m$ could be placed on either side of
the propagator. In this derivation, we chose to put it to the left
of the propagator in the first term in Eq.~(\ref{spectral density
time derivative}) and to the right of the propagator in the second
term.  If instead one places $\hat{H}_m$ to the right of the
propagator in the first term and to the left of the propagator in
the second term we can obtain an alternate form of the equation of
motion\footnote{If $\hat{H}_m$ is placed in other locations, say,
either to the right or left in both terms, the evolution operator
for $W$ is the mean of the two forms discussed in the text.}. The
manipulations are similar to those described above and in the
Appendix and are not repeated her. The resulting equation of
motion is
\begin{eqnarray}\label{spectral density time derivative7}
\frac{\partial W(t)}{\partial t} &=& + \frac{1}{\hbar}
\sum_{\lambda \lambda'} h_{\lambda
\lambda'}(R_2)\Big[r_{2\lambda}\frac{\partial}{\partial
p_{2\lambda'}} - p_{2 \lambda} \frac{\partial}{\partial
r_{2\lambda'}} \Big] W(t) -(\frac{P_2}{M} \cdot \frac{\partial }{\partial
R_2} - \frac{\partial
H_m}{\partial R_2} \cdot \frac{\partial }{\partial P_2})W(t)  \nonumber \\
&&- \frac{\hbar}{8} \sum_{\lambda \lambda'} \frac{\partial
h_{\lambda \lambda'}(R_2)}{\partial R_2}
\cdot \Big(\frac{\partial}{\partial
r_{2\lambda}}\frac{\partial}{\partial r_{2\lambda'}} +
\frac{\partial}{\partial p_{2\lambda'}}\frac{\partial}{\partial
p_{2\lambda}}\Big) \frac{\partial}{\partial P_2} W(t)\nonumber \\
&\equiv& -i{\mathcal L}_m(x_2,X_2) W(t).
\end{eqnarray}
This alternate form of the dynamics not only shows the symmetry of
the dynamics of W in its variables but also allows us to move the
time evolution from W to the observable $B_{Wm}$ in the
correlation function expression. The formal solution of
Eq.~(\ref{spectral density time derivative7}) is
\begin{equation} \label{eq:formal-soln}
W(X_1,X_2,x_1,x_2,t)=e^{-i{\mathcal L}_m(x_2,X_2) t}
W(X_1,X_2,x_1,x_2,0).
\end{equation}
These results will be used in the next section to derive a
quantum-classical approximation to the time correlation function.

\section{Quantum-Classical Correlation Function} \label{sec:QC-corr-function}
We can now employ the results of the last two sections to find an
expression for the quantum-classical approximation to the
correlation function in the mapping basis. Using
Eq.~(\ref{eq:formal-soln}) in Eq.~(\ref{correlation5a}), we have
\begin{eqnarray} \label{QC-correlation}
C_{AB}(t) &=& \frac{1}{(2\pi\hbar)^{(N_b+N)}}
\int dX_1dX_2 dx_1 dx_2\; ({\hat \rho}_{eq}{\hat A})_{Wm}(X_1,x_1) \nonumber \\
&&\times {B}_{Wm}(X_2,x_2) e^{-i{\mathcal L}_m(x_2,X_2) t}
W(X_1,X_2,x_1,x_2,0).
\end{eqnarray}
Performing an integration by parts to move the evolution operator onto the
${B}_{Wm}$ and integrating over the coordinates with subscript 1,
making use of the delta functions in the initial value of $W$, yields,
\begin{eqnarray} \label{QC-correlationb}
C_{AB}(t) &=& \int dX dx\;
({\hat \rho}_{eq}{\hat A})_{Wm}(X,x) {B}_{Wm}(X,x,t),
\end{eqnarray}
where ${B}_{Wm}(X,x,t)=e^{i{\mathcal L}_m(x,X) t}{B}_{Wm}(X,x)$.
(We have dropped the subscripts on the phase space coordinates
since this notation is no longer needed.) This expression contains
the full quantum equilibrium structure of the subsystem and bath
and the quantum-classical Liouville evolution of the operator
$\hat{B}$ in the mapping representation.

The evolution equation for ${B}_{Wm}(X,x,t)$ that one obtains in
this derivation is identical to that found earlier in the
calculation of the average value of an observable~\cite{hyo:ray:ali}. This evolution
equation can be written in the form
\begin{eqnarray}\label{eq:map-obs}
 &  & \frac{d}{dt}B_{Wm}(x,X,t)= i{\mathcal{L}}_{m}B_{Wm}(t)= -\{H_{m},B_{Wm}(t)\}_{x,X}\label{eq:awt3}\\
 &  & \quad+\frac{\hbar}{8}\sum_{\lambda\lambda'}\frac{\partial h_{\lambda\lambda'}}
 {\partial R}\cdot(\frac{\partial}{\partial r_{\lambda'}}\frac{\partial}{\partial r_{\lambda}}
 +\frac{\partial}{\partial p_{\lambda'}}\frac{\partial}{\partial p_{\lambda}})
 \frac{\partial}{\partial P}B_{Wm}(t),\nonumber
 \end{eqnarray}
 where $\{\;,\;\}_{x,X}$ denotes a Poisson bracket in the
full mapping-bath phase space of the system. The quantum-classical
Liouville operator in the mapping basis can be decomposed into two
terms,
$i{\mathcal{L}}_{m}=i{\mathcal{L}}_{m}^{0}+i{\mathcal{L}}_{m}^{\prime}$
where \begin{eqnarray}
i{\mathcal{L}}_{m}^{0} & = & -\{H_{m},\quad \}_{x,X},\\
i{\mathcal{L}}_{m}^{\prime} & = &
\frac{\hbar}{8}\sum_{\lambda\lambda'}\frac{\partial
h_{\lambda\lambda'}}{\partial R}\cdot(\frac{\partial}{\partial
r_{\lambda'}}\frac{\partial}{\partial
r_{\lambda}}+\frac{\partial}{\partial
p_{\lambda'}}\frac{\partial}{\partial
p_{\lambda}})\frac{\partial}{\partial P}.\nonumber
\end{eqnarray}
The $i{\mathcal{L}}_{m}^{0}$ evolution operator, corresponding to
the Poisson bracket in Eq.~(\ref{eq:awt3}), leads to a
classical-like evolution of the coupled dynamics of the quantum
mapping and classical bath phase space variables that can be
simulated by Newtonian trajectories. The force field that the
classical variables feel varies with time as a result of the
dependence of the forces on the mapping phase space variables. If
$i{\mathcal{L}}_{m}^{\prime}$ is dropped in the evolution
equation, $B_{Wm}(t)$ has a solution in terms of characteristics.
The set of ordinary differential equations that determines its
solution is
\begin{eqnarray}\label{eq:class}
\frac{dr_{\lambda}(t)}{dt} & = & \frac{1}{\hbar}\sum_{\lambda'}h_{\lambda\lambda'}(R(t))p_{\lambda'}(t),\nonumber \\
\frac{dp_{\lambda}(t)}{dt} & = & -\frac{1}{\hbar}\sum_{\lambda'}h_{\lambda\lambda'}(R(t))r_{\lambda'}(t),\nonumber \\
\frac{dR(t)}{dt} & = &
\frac{P(t)}{M},\quad\frac{dP(t)}{dt}=-\frac{\partial
H_{m}}{\partial R(t)}.
\end{eqnarray}
The utility of this approximation to Eq.~(\ref{eq:map-obs}) is a
topic of current research. Tests of its accuracy have been carried
out on the spin-boson model where Eq.~(\ref{eq:map-obs}) is
equivalent to full quantum dynamics and are being carried out on
other model systems with nonlinear coupling between the quantum
and classical degrees of freedom where Eq.~(\ref{eq:map-obs}) is
not exact. For the spin-boson model Eq.~(\ref{eq:class}) yields
results that are indistinguishable from the known exact quantum
results for this system~\cite{hyo:ray:ali}. Tests being carried
out on other model systems show that while the results are often
in close accord with exact quantum results, sometimes
discrepancies are observed that cannot be ascribed to
approximations in Eq.~(\ref{eq:map-obs}) for these more general
interactions and must be attributed to the use of
Eq.~(\ref{eq:class}) for the dynamics. Consequently, further
research is underway to fully characterize the contributions
arising from $i{\mathcal{L}}_{m}^{\prime}$ and construct
algorithms that account for its action.

\section{Conclusion} \label{sec:conc}

In the adiabatic basis the correlation function may be evaluated
in terms of an ensemble of surface-hopping trajectories~\cite{ray99, Sergi:MacKernan}. While
this simulation scheme has been used in a number of applications
it is difficult to obtain accurate results for long times due the
presence of oscillating terms in the Monte Carlo sampling. In this
paper, we show that it is possible to reformulate the calculation of
the correlation function in the mapping basis.
Simulation schemes based on this formulation do not suffer from
some of the problems that arise in the implementation of
surface-hopping schemes.

In order to evaluate the expression for the correlation function
obtained in this paper two ingredients are required. The
expression involves an average over the full quantum equilibrium
structure. As briefly discussed above, the computation of quantum
equilibrium structure is a more tractable problem than the
calculation of quantum dynamics. Nevertheless, approximations are
usually required to evaluate the equilibrium structure. In the
high temperature limit it is possible to derive analytical
expressions that are useful in many applications~\cite{Kim:Hanna, Kim:Kapral, Hanna:Kapral}.

The main result of this paper is the quantum-classical expression
for the correlation function that involves time evolution of the
quantum subsystem in the mapping basis. Provided the
quantum-classical evolution is approximated by $i{\mathcal{L}}_{m}
\simeq i{\mathcal{L}}_{m}^{0}$ the time evolution of the dynamical
variable in the correlation function can be computed easily by
solving a set of Newtonian-like equations. Thus, difficulties
associated with the accumulation of Monte Carlo weights in the
evaluation of an oscillatory function that arise in the
surface-hopping solution of the quantum-classical Liouville
equation in the adiabatic basis are by-passed. Of course, this
simple scheme relies on the ability to neglect
$i{\mathcal{L}}_{m}^{\prime}$, which accounts for higher order
correlations in the dynamics. Thus, the focus of future research
is on the characterization of the nature of the dynamics generated
by $i{\mathcal{L}}_{m}^{\prime}$ and the construction of
simulation algorithms that account for its presence. The results
in this paper form the basis for future applications to the
calculation of transport properties, such as rate constants for
nonadiabtic chemical reactions.

\noindent {\it Acknowledgement}: This work was in part supported
by a grant from the Natural Science and Engineering Research
Council of Canada. AN acknowledges the support from the Lachlan Gilchrist fellowship.
\appendix

\section*{Appendix: Derivation of the Dynamics}
In this Appendix, we give the details of the calculations needed to
obtain the quantum-classical evolution equation for the spectral
density function. There are ten terms in Eq.~(\ref{spectral
density time derivative2}) and, because of the symmetry of the
expression, it is convenient to group term i with term i+5 and
evaluate the contributions group by group. First, we perform the
integrals over $Q$ and $q$. The 1st and 6th terms contain
$\frac{\partial^2}{\partial \tilde Z_1^2}$ and
$\frac{\partial^2}{\partial \tilde Z_2^2}$, respectively, after this integration.
Integration by parts with respect to $\tilde Z_1$ and $\tilde Z_2$ and
summation of the results yield $\mu P_1 \cdot \frac{\partial
W}{\partial R_1}$. The 2nd and 7th terms involve bath potentials.
After expansion of $V_B(R_1 \pm \frac{\mu \tilde Z_1}{2})$ in $\mu$, the first and third terms in the series, which are
proportional to $\mu^0$ and $\mu^2$, cancel and the second terms, which are
proportional to $\mu$, yield $- \mu \frac{\partial
V_B(R_1)}{\partial R_1} \cdot \frac{\partial W}{\partial P_1}$. The contribution from the 5th and 10th terms derived similarly to
yield $\frac{\mu}{2}\frac{\partial \rm{Tr}(h)}{\partial R_1} \cdot
\frac{\partial W}{\partial P_1}$.
The derivations of the third and eighth and, also, forth and fifth groups of terms are more complicated
and are presented in the following two subsections.

\subsection*{\it 3rd and 8th terms:}
The sum of the 3rd and 8th contributions to the time derivative of
$W$ is
\begin{eqnarray}
\label{3rd&8st}
&&\Big(\frac{\partial W}{\partial t}\Big)_{3+8} = \frac{i\mu^{2N_b}}{2} \sum_{\lambda \lambda'} \int d \tilde Z_1 d \tilde Z_2 dz_1 dz_2 e^{-i(P_1 \cdot \tilde Z_1+P_2 \cdot \tilde Z_2+ p_1 \cdot z_1+p_2\cdot z_2)}\nonumber \\
&&\qquad \times \Bigg \{ \Big[h_{\lambda \lambda'}(R_1)+\frac{\partial h_{\lambda \lambda'}(R_1)}
{\partial R_1} \cdot \frac{\mu \tilde Z_1}{2}
\Big] \Big[r_{1\lambda}r_{1\lambda'}+r_{1\lambda}z_{1\lambda'}/2
+z_{1\lambda}r_{1\lambda'}/2+z_{1\lambda}z_{1\lambda'}/4 \Big] \nonumber \\
&&\qquad - \Big[h_{\lambda \lambda'}(R_1)-\frac{\partial h_{\lambda \lambda'}(R_1)}{\partial R_1} \cdot
\frac{\mu \tilde Z_1}{2}
\Big] \Big[r_{1\lambda}r_{1\lambda'}-r_{1\lambda}z_{1\lambda'}/2-z_{1\lambda}r_{1\lambda'}/2
+z_{1\lambda}z_{1\lambda'}/4 \Big] \Bigg \} \nonumber \\
&&\qquad \times \langle r_1+\frac{z_1}{2}|\langle R_1+\frac{\mu \tilde Z_1}{2}|e^{i\hat{H}_mt}|R_2-\frac{\mu \tilde Z_2}{2} \rangle
|r_2-\frac{z_2}{2} \rangle \nonumber \\
&&\qquad \times \langle r_2+\frac{z_2}{2}|\langle
R_2+\frac{\mu \tilde Z_2}{2}|e^{-i\hat{H}_mt}|R_1-\frac{\mu \tilde Z_1}{2}
\rangle |r_1-\frac{z_1}{2} \rangle, \nonumber
\end{eqnarray}
where we have performed a McLauren expansion of
$h_{\lambda \lambda'}$ in the small parameter $\mu$, and retained the first two terms. Finally, using the fact that a partial
differential with respect to momentum acting on the exponential
term has the same effect as multiplication by the variable $z_1$ or $\tilde Z_1$
that results from the expansion, we have
\begin{eqnarray}
\label{3rd&8s} \Big(\frac{\partial W}{\partial t}\Big)_{3+8} &=&\nonumber
 \frac{i\mu^{2N_b}}{2} \sum_{\lambda \lambda'} \int d \tilde Z_1 d \tilde Z_2 dz_1 dz_2 \Big \{ i h_{\lambda \lambda'}(R_1)
 \big [r_{1\lambda}\frac{\partial}{\partial p_{1\lambda'}}+r_{1\lambda'}\frac{\partial}{\partial p_{1\lambda}}\big ]  \\
&&+ \frac{\mu}{2} \frac{\partial h_{\lambda \lambda'}(R_1)}{\partial R_1} \cdot(i\frac{\partial}{\partial P_{1}})
\big [2r_{1\lambda}r_{1\lambda'} - \frac{1}{2} \frac{\partial}{\partial p_{1\lambda}}
\frac{\partial}{\partial p_{1\lambda'}}\big ]
 \Big \} e^{-i(P_1 \cdot \tilde Z_1+P_2 \cdot \tilde Z_2+p_1 \cdot z_1 + p_2 \cdot z_2)} \nonumber \\
&&\times \langle r_1+\frac{z_1}{2}|\langle R_1+\frac{\mu \tilde Z_1}{2}|
e^{i\hat{H}_mt}|R_2-\frac{\mu \tilde Z_2}{2} \rangle |r_2-\frac{z_2}{2}
\rangle \nonumber \\
&&\times \langle r_2+\frac{z_2}{2}|\langle R_2+\frac{\mu \tilde Z_2}{2}|
e^{-i\hat{H}_mt}|R_1-\frac{\mu \tilde Z_1}{2} \rangle |r_1-\frac{z_1}{2} \rangle  \nonumber \\
&=& -\frac{1}{2} \sum_{\lambda \lambda'} h_{\lambda
\lambda'}(R_1)\big [r_{1\lambda}\frac{\partial}{\partial
p_{1\lambda'}}+r_{1\lambda'}\frac{\partial}{\partial
p_{1\lambda}}\big ] W \nonumber \\
&&- \frac{\mu}{4} \sum_{\lambda \lambda'}
\frac{\partial h_{\lambda \lambda'}(R_1)}{\partial
R_1} \cdot \big [2r_{1\lambda}r_{1\lambda'} -\frac{1}{2}
\frac{\partial}{\partial p_{1\lambda}}\frac{\partial}{\partial
p_{1\lambda'}}\big ] \frac{\partial}{\partial P_{1}} W, \nonumber
\end{eqnarray}
where in the last equality the expression for the scaled W is inserted.

\subsection*{\it 4th and 9th terms:}
The sum of the 4th and 9th terms in Eq.~(\ref{spectral density time derivative2}) is
\begin{eqnarray}
\label{4st&9st}
&&\Big(\frac{\partial W}{\partial t}\Big)_{4+9} = -\frac{i\mu^{2N_b}}{2}\int
d \tilde Z_1 d \tilde Z_2 dz_1 dz_2
\sum_{\lambda \lambda'}\Bigg \{ \Big [ h_{\lambda \lambda'}(R_1+\frac{\mu \tilde Z_1}{2})
\frac{\partial}{\partial (r_1+\frac{z_1}{2})_{\lambda}}
\frac{\partial}{\partial (r_1+\frac{z_1}{2})_{\lambda'}} \nonumber \\
&&\qquad \times \langle r_1+\frac{z_1}{2}|\langle R_1+\frac{\mu \tilde Z_1}{2}|
e^{i\hat{H}_mt}|R_2-\frac{\mu \tilde Z_2}{2} \rangle |r_2-\frac{z_2}{2}
\rangle \Big ]  \nonumber \\
&&\qquad \times \langle r_2+\frac{z_2}{2}|
\langle R_2+\frac{\mu \tilde Z_2}{2}|e^{-i\hat{H}_mt}|R_1-\frac{\mu \tilde Z_1}{2}
\rangle |r_1-\frac{z_1}{2} \rangle \nonumber \\
&&\qquad - \Big[ h_{\lambda \lambda'}(R_1-\frac{\mu \tilde Z_1}{2})
\frac{\partial}{\partial (r_1-\frac{z_1}{2})_{\lambda}}
\frac{\partial}{\partial (r_1-\frac{z_1}{2})_{\lambda'}} \nonumber \\
&& \qquad \times \langle r_2+\frac{z_2}{2}|\langle R_2+\frac{\mu \tilde Z_2}{2}|
e^{-i\hat{H}_mt}|R_1-\frac{\mu \tilde Z_1}{2}>|r_1-\frac{z_1}{2} \rangle \Big]
\nonumber \\
&&\qquad \times \langle r_1+\frac{z_1}{2}|\langle R_1+\frac{\mu \tilde Z_1}{2}|
e^{i\hat{H}_mt}|R_2-\frac{\mu \tilde Z_2}{2} \rangle |r_2-\frac{z_2}{2}
\rangle  \Bigg \} e^{-i(P_1 \cdot \tilde Z_1+P_2 \cdot \tilde Z_2+ p_1
\cdot z_1+p_2\cdot z_2)} \nonumber \\
&& \qquad \qquad \quad = -\frac{i\mu^{2N_b}}{2} \sum_{\lambda \lambda'} \int d
\tilde Z_1 d \tilde Z_2 dz_1 dz_2 \Bigg\{ \Big[\big(h_{\lambda \lambda'}(R_1)+\frac{\partial
h_{\lambda \lambda'}(R_1)}{\partial R_1}\frac{\mu \tilde Z_1}{2}
\big) \nonumber \\
&& \qquad \times \frac{\partial}{\partial r_{1\lambda}}\frac{\partial}{\partial
r_{1\lambda'}} \langle r_1+\frac{z_1}{2}|\langle R_1+\frac{\mu \tilde Z_1}{2}|
e^{i\hat{H}_mt}|R_2-\frac{\mu \tilde Z_2}{2} \rangle|r_2-\frac{z_2}{2} \rangle \Big]
\nonumber  \\
&&\qquad \times \langle r_2+\frac{z_2}{2}|\langle R_2+\frac{\mu \tilde Z_2}{2}|
e^{-i\hat{H}_mt}|R_1-\frac{\mu \tilde Z_1}{2}
 \rangle |r_1-\frac{z_1}{2} \rangle  \nonumber \\
&&\qquad  - \Big[\big(h_{\lambda \lambda'}(R_1)-
\frac{\partial h_{\lambda \lambda'}(R_1)}{\partial R_1}\frac{\mu \tilde Z_1}{2}
 \big)\nonumber \\
&&\qquad \times\frac{\partial}{\partial r_{1\lambda}}\frac{\partial}{\partial r_{1\lambda'}} \langle r_2+\frac{z_2}{2}|
\langle R_2+\frac{\mu \tilde Z_2}{2}|e^{-i\hat{H}_mt}|R_1-\frac{\mu \tilde Z_1}{2}
\rangle  |r_1-\frac{z_1}{2} \rangle \Big] \nonumber \\
&&\qquad \times \langle r_1+\frac{z_1}{2}|\langle R_1+\frac{\mu \tilde Z_1}{2}|
e^{i\hat{H}_mt}|R_2-\frac{\mu \tilde Z_2}{2} \rangle  |r_2-\frac{z_2}{2} \rangle  \Bigg\}e^{-i(P_1 \cdot \tilde Z_1+P_2 \cdot \tilde Z_2+p_1 \cdot z_1+p_2 \cdot z_2)}.
\end{eqnarray}
Again in the second equality we have carried out a McLauren
expansion of $h_{\lambda \lambda'}$ to first order in $\mu$.
Equation~(\ref{4st&9st}) itself has four terms. In the sum of the
1st and 3rd subcontributions one of the partial derivatives over $r$
may be replaced by a partial derivative over $z$ and an integration
by parts may be carried out. We find
\begin{eqnarray}
\label{1st&3rd}
&&\frac{i\mu^{2N_b}}{2} \sum_{\lambda \lambda'} h_{\lambda \lambda'}(R_1) \int d \tilde Z_1 d \tilde Z_2 dz_1 dz_2
e^{-i(P_1 \cdot \tilde Z_1+P_2 \cdot \tilde Z_2+p_1 \cdot z_1+p_2 \cdot z_2)} \nonumber \\
&&\quad \times \Bigg \{ \Big [2\frac{\partial}{\partial r_{1\lambda'}}
\langle r_1+\frac{z_1}{2}|
\langle R_1+\frac{\mu \tilde Z_1}{2}|e^{i\hat{H}_mt}|R_2-\frac{\mu \tilde Z_2}{2} \rangle  |r_2-\frac{z_2}{2} \rangle  \nonumber \\
&&\quad \times (\frac{\partial}{\partial z_{1\lambda}} -
ip_{1\lambda})  \langle r_2+\frac{z_2}{2}|
\langle R_2+\frac{\mu \tilde Z_2}{2}|e^{-i\hat{H}_mt}|R_1-\frac{\mu \tilde Z_1}{2} \rangle  |r_1-\frac{z_1}{2} \rangle \Big ]\nonumber  \\
&&\quad - \Big [ (-2)\frac{\partial}{\partial r_{1\lambda'}} \langle
r_2+\frac{z_2}{2}|\langle R_2+\frac{\mu \tilde Z_2}{2}|
e^{-i\hat{H}_mt}|R_1-\frac{\mu \tilde Z_1}{2} \rangle  |r_1-\frac{z_1}{2} \rangle   \nonumber \\
&&\quad \times (\frac{\partial}{\partial z_{1\lambda}} -
ip_{1\lambda}) \langle r_1+\frac{z_1}{2}|
\langle R_1+\frac{\mu \tilde Z_1}{2}|e^{i\hat{H}_mt}|R_2-\frac{\mu \tilde Z_2}{2} \rangle  |r_2-\frac{z_2}{2} \rangle \Big ] \Bigg \} \nonumber \\
&&= \sum_{\lambda \lambda'} h_{\lambda \lambda'}(R_1) p_{1\lambda}
\frac{\partial}{\partial r_{1\lambda'}} W.
\end{eqnarray}
Here we see that the two terms with a momentum multiplier are
added together while the other terms cancel each other.

The 2nd and 4th subcontributions in Eq.~(\ref{4st&9st}) are treated in the
following way: We replace multiplication by the variable $\tilde Z$ with a partial
differentiation with respect to the momentum.
Furthermore, the sum of these contributions is written as one half the
sum of two equal contributions, the expressions appearing in Eq. (\ref{4st&9st}) and
the same expression but with the differentials over $r$ replaced with those over $z$.
Thus, the sum of the 2nd and 4th contributions can be written as
\begin{eqnarray}
\label{2nd&4st}
&& -\frac{i\mu^{2N_b+1}}{8} \sum_{\lambda \lambda'} \frac{\partial h_{\lambda \lambda'}(R_1)}{\partial R_1}
\int d \tilde Z_1 d \tilde Z_2 dz_1 dz_2 (i\frac{\partial}{\partial P_1}) e^{-i(P_1 \cdot \tilde Z_1+P_2 \cdot \tilde Z_2 + p_1 \cdot z_1+p_2 \cdot z_2)}\nonumber \\
&&\times \Bigg\{ \bigg[ \Big ( \frac{\partial}{\partial r_{1\lambda}}\frac{\partial}{\partial r_{1\lambda'}} \langle
r_1+\frac{z_1}{2}|\langle R_1+\frac{\mu \tilde Z_1}{2}|e^{i\hat{H}_mt}|R_2-\frac{\mu \tilde Z_2}{2}
 \rangle  |r_2-\frac{z_2}{2} \rangle \Big )\nonumber  \\
&&\times \langle r_2+\frac{z_2}{2}|\langle R_2+\frac{\mu \tilde Z_2}{2}|e^{-i\hat{H}_mt}|R_1-\frac{\mu \tilde Z_1}{2}
 \rangle  |r_1-\frac{z_1}{2} \rangle   \nonumber \\
&&+  \langle r_1+\frac{z_1}{2}|\langle R_1+\frac{\mu \tilde Z_1}{2}|e^{i\hat{H}_mt}|R_2-\frac{\mu \tilde Z_2}{2}
 \rangle  |r_2-\frac{z_2}{2} \rangle   \nonumber \\
&&\times \Big (\frac{\partial}{\partial r_{1\lambda}}\frac{\partial}{\partial r_{1\lambda'}}
\langle r_2+\frac{z_2}{2}|\langle R_2+\frac{\mu \tilde Z_2}{2}|
e^{-i\hat{H}_mt}|R_1-\frac{\mu \tilde Z_1}{2} \rangle  |r_1-\frac{z_1}{2} \rangle \Big ) \bigg] \nonumber \\
&&+ 4 \bigg[\Big (\frac{\partial}{\partial z_{1\lambda}}\frac{\partial}{\partial z_{1\lambda'}}
\langle r_1+\frac{z_1}{2}|\langle R_1+\frac{\mu  \tilde Z_1}{2}|
e^{i\hat{H}_mt}|R_2-\frac{\mu \tilde  Z_2}{2} \rangle  |r_2-\frac{z_2}{2} \rangle \Big ) \nonumber \\
&&\times \langle r_2+\frac{z_2}{2}|\langle R_2+\frac{\mu \tilde
Z_2}{2}|e^{-i\hat{H}_mt}|R_1-\frac{\mu \tilde  Z_1}{2} \rangle  |r_1-\frac{z_1}{2} \rangle  \nonumber  \\
&&+ \langle r_1+\frac{z_1}{2}|\langle R_1+\frac{\mu \tilde Z_1}{2}|e^{i\hat{H}_mt}|R_2-\frac{\mu \tilde  Z_2}{2}
 \rangle  |r_2-\frac{z_2}{2} \rangle  \nonumber  \\
&& \times \Big (\frac{\partial}{\partial z_{1\lambda}}\frac{\partial}{\partial z_{1\lambda'}} \langle
r_2+\frac{z_2}{2}|\langle R_2+\frac{\mu \tilde Z_2}{2}|e^{-i\hat{H}_mt}|R_1-\frac{\mu \tilde  Z_1}{2} \rangle
|r_1-\frac{z_1}{2} \rangle \Big ) \bigg] \Bigg \}.
\end{eqnarray}
 If in both expressions we replace $\frac{\partial^2 A}{\partial x \partial y} B+ A\frac{\partial^2 B}{\partial x \partial y}$ with
 $\frac{\partial^2 (AB)}{\partial x \partial y} - \frac{\partial A}{\partial x}\frac{\partial B}{\partial y} - \frac{\partial A}{\partial y}\frac{\partial B}{\partial x}$,
 the cross terms cancel. Finally, integrating by parts over
 $z_{1\lambda}$ and $z_{1\lambda'}$, we get
\begin{equation}
\label{2nd&4sta}
\frac{\mu}{8}  \sum_{\lambda \lambda'} \frac{\partial
h_{\lambda \lambda'}(R_1)}{\partial R_1} \cdot (\frac{\partial}{\partial
r_{1\lambda}}\frac{\partial}{\partial r_{1\lambda'}} -4
p_{1\lambda}p_{1\lambda'}) \frac{\partial}{\partial P_1} W.
\end{equation}
Summing all the contributions from the above calculations, we find
the comparatively simple result
\begin{eqnarray}
\label{spectral density time derivative3} \frac{\partial
W}{\partial t} &=& - \sum_{\lambda \lambda'} h_{\lambda
\lambda'}(R_1) \Big[r_{1\lambda}\frac{\partial}{\partial
p_{1\lambda'}} - p_{1\lambda} \frac{\partial}{\partial
r_{1\lambda'}} \Big]W\nonumber \\
&&+ \mu P_1 \cdot \frac{\partial W}{\partial R_1} - \mu \frac{\partial
V_B(R_1)}{\partial R_1} \cdot \frac{\partial W}{\partial P_1}  -
\frac{\mu}{2} \sum_{\lambda \lambda'} \frac{\partial h_{\lambda
\lambda'}(R_1)}{\partial R_1} \cdot (r_{1\lambda}r_{1\lambda'} +
p_{1\lambda}p_{1\lambda'} - \delta_{\lambda
\lambda'})\frac{\partial}{\partial P_{1}} W\nonumber \\
&& + \frac{\mu}{8}  \sum_{\lambda \lambda'} \frac{\partial
h_{\lambda \lambda'}(R_1)}{\partial R_1} \cdot
\Big(\frac{\partial}{\partial
r_{1\lambda}}\frac{\partial}{\partial r_{1\lambda'}} +
\frac{\partial}{\partial p_{1\lambda}}\frac{\partial}{\partial
p_{1\lambda'}}\Big) \frac{\partial}{\partial P_1} W .
\end{eqnarray}
Restoring unscaled coordinates we have Eq.~(\ref{spectral density time derivative5}) in the text.
\bibliography{correlation2}
\end{document}